\documentclass[twocolumn,showpacs,prl,aps,amssymb]{revtex4}
\usepackage{graphicx}
\begin{document}

\title{\bf

A simple test for hidden variables in  spin-1 system

}

\author{Alexander A. Klyachko,  M. Ali Can, Sinem Binicio\u{g}lu, and Alexander S. Shumovsky}

\affiliation{Faculty of Science, Bilkent University, Bilkent,
Ankara, 06800 Turkey}

\begin{abstract}
We resolve an old problem about the existence of hidden parameters
in a three-dimensional quantum system by constructing an
appropriate Bell's type inequality. This reveals a nonclassical
nature of most spin-$1$ states. We shortly discuss some physical
implications and an underlying cause of this nonclassical
behavior, as well as a perspective of its experimental
verification.

\end{abstract}

\pacs{03.65.Ud, 03.65.-w, 03.65.Ta}

\maketitle

The concept of {\it quantum entanglement}, as well as a prospect
of its applications in quantum computing, have attracted a great
deal of interest \cite{all}. No doubt, its most striking
manifestation is {\it quantum nonlocality\/}, understood here as a
correlation beyond light cones of spatially separated quantum
systems, where no classical interaction between them is possible.
However, for quantum computation a magic ability of entanglement
to bypass constraints imposed by the so called {\it classical
realism\/} is far more important. The latter is understood here as
the existence of hidden parameters, or equivalently a joint
probability distribution of all involved quantum observables. This
property of entanglement can be modelled by no classical device,
which emphasizes  a  qualitative distinction between classical and
quantum information processing, beyond a mere difference   in
their computational power.

Following Bell's seminal  works \cite{Bell}, the nonclassical
behavior is usually detected by violation of certain inequalities,
collectively named Bell's conditions. Their experimental test
\cite{Aspect} left  little or no doubt that entangled states
indeed override the  classical constraints, in spite of
everlasting search for possible loopholes
\cite{
review}.

Initially Bell justified his constraints by {\it locality
argument\/} that space-like separated quantum systems  can not
affect each other. This approach neither  excludes the existence
of {\it nonlocal\/} hidden parameters nor can be applied to local
systems. In this letter we provide a test for hidden variables in
{\it local\/} spin-$1$ system, where the original Bell's approach
clearly failes. We found that every spin state is nonclassical,
except for {\it coherent\/} one $|1\rangle_\ell$ with spin
projection $1$ onto some direction $\ell$.

To elucidate the physical  difference between the coherent state
$|1\rangle$ and its antithetic counterpart $|0\rangle$, called
{\it neutrally polarized\/} spin state, consider a decay of spin-1
system into two spin-$1/2$ components. The resulting two particle
state must be symmetric under the exchange of the components and
preserve the angular momentum. As a result, the coherent state
$|1\rangle$ decays  into the separable one
$\mid\uparrow\uparrow\rangle$, while the neutrally polarized state
$|0\rangle$  yields  the maximally entangled Bell state
$\frac{1}{\sqrt{2}}(\mid\uparrow\downarrow\rangle+\mid\downarrow\uparrow\rangle)$.

The problem we address here is whether one can detect something
non-classical in the state $|0\rangle$ before the decay? Recall
that by the Kochen--Specker theorem \cite{KS} every spin-$1$ state
is incompatible with the so-called {\it context-free\/} hidden
variables model. The latter entertains the notion of ``hidden
value" of an observable $A$ revealed by its measurement and
independent of a measurement of any other observable $B$ commuting
with $A$. Bell found no physical ground for this hypothesis and
eventually abandoned it in favor of inequalities based on
locality rather than noncontextuality. However, on the way he
switched from spin-$1$ system to system of two qubits, leaving the
problem of existence of {\it contextual\/} hidden parameters in
spin-1 system open. We resolve this problem below.

Our approach to hidden variables in spin systems is similar to
that of Fine \cite{Fine} for two qubits  with two measurements
$A_1,A_2$ and $B_1,B_2$ at sites $A$ and $B$ respectively. The
hidden variables provide a joint probability distribution of all
four observables $A_1,A_2,B_1,B_2$ compatible with the
distributions of   commuting pairs $A_i,B_j$ predicted by quantum
mechanics and available for experimental verification. The arising
general problem of existence of a joint probability distribution
of random variables $x_1,x_2,\ldots,x_n$, compatible with given
partial distributions of some of them, is known as the {\it
marginal problem\/} \cite{MP_Appl}. Geometrically it amounts to
existence of a ``body" in $\mathbb{R}^n$ of a nonnegative density
with given projections onto some coordinate subspaces.

The problem has been settled in early sixties \cite{MargProb}.
Applying the solution to observables $A_i,i=1,2,\ldots,n$ in an
arbitrary finite quantum system we arrive at the following {\it
ansatz\/} for testing classical realism
\cite[$n^\circ$2.3]{Klyachko02}. Let $a_i$ be a variable
describing all possible outcomes of the observable $A_i$. We will
use the shortcuts $A_I$ for the subset of observables $A_i, i\in
I$, and $a_I$ for the corresponding subset of variables $a_i, i\in
I$. Consider now a non-negative function of the form
\begin{equation}\label{Kell_Cone}
\sum_{A_I\text{ commute}} f_I(a_I)\ge 0
\end{equation}
and assume the existence of a hidden distribution of all variable
$a_1,a_2,\ldots,a_n$ compatible with the distributions of commuting
observables $a_I$ predicted by quantum mechanics. Then, taking
expectation value of Eq.~(\ref{Kell_Cone}) with respect to the
hidden distribution, we arrive at the Bell type inequality for
testing classical realism
\begin{equation}\label{ansatz}
\sum_{A_I\text{ commute}} \langle\psi|f_I(A_I)|\psi\rangle\ge 0.
\end{equation}
 It turns out that these
inequalities are also {\it sufficient\/} for the existence of
hidden variables \cite{MargProb}. To make this criterion
effective, observe that the set of all functions given by
Eq.~(\ref{Kell_Cone}) form a polyhedral cone, called {\it
Vorob'ev--Kellerer cone\/}, and the conditions (\ref{ansatz})
should be checked for its {\it extremal edges\/} only. The latter
can be routinely found using an appropriate software, e.g.
\texttt{Convex} package \cite{Franz}.

As a result, we end up with a {\it finite\/} set of inequalities
that are necessary and  sufficient for an extension of the partial
distributions of commuting observables $A_I$ to a hidden
distribution of all observables $A_i$, commuting or not. The
latter can be modelled by classical means like tossing dice. This
makes the quantum system indistinguishable from a classical one.

Let's separate the extremal edges generated by a single function
$f_I(a_I)\ge0$ vanishing everywhere except one point. The
corresponding Bell inequality is vacuous and we call such extremal
functions {\it trivial\/}.

For two qubits the {\it ansatz\/}  gives 8 nontrivial extremal
functions. The respective constraints can be obtained from
Clauser-Horne-Shimoni-Holt inequality \cite{B-conditions}
\begin{equation}\label{CHSH}
\langle A_1B_1\rangle+\langle
A_1B_2\rangle+\langle A_2B_1\rangle-\langle A_2B_2\rangle
\le2,
\end{equation}
by spin flips $A_i\mapsto \pm A_i,B_j\mapsto
\pm B_j$.
This criterion for existence of hidden parameters 
was first proved by Fine~\cite{Fine}.

Returning to spin-1 system, consider a cyclic quintuplet  of unit
vectors $\ell_i\perp\ell_{i+1}$ with the indices taken modulo 5,
see Fig~1. We call it a {\it pentagram\/}. The orthogonality
implies that squares of spin projection operators $S_{\ell_i}$
onto directions $\ell_i$ commute for successive indices
$[S_{\ell_i}^2,S_{\ell_{i+1}}^2]=0$. We find it more convenient to
deal with the observables $A_i=2S_{\ell_i}^2-1$ taking values
$a_i=\pm1$. They satisfy the following inequality
\begin{eqnarray}
a_1a_2+a_2a_3+a_3a_4+a_4a_5+a_5a_1+3 \geq 0. \label{inequality-1}
\end{eqnarray}
Indeed, the product of the monomials in the left hand side is
equal to $1$, hence at least one term is equal to $+1$, and the
sum of the rest is no less than $-4$.

Assuming now the existence of a hidden distribution of all
observables $a_i$, and taking the respective  expectation value of
Eq.~(\ref{inequality-1}) we arrive at the inequality
\begin{equation}\label{reflect}
\langle A_1A_2\rangle+\langle A_2A_3\rangle+\langle A_3A_4\rangle+
\langle A_4A_5\rangle+\langle A_5A_1\rangle\geq -3,
\end{equation}
that can be recast into the form
\begin{equation}\label{kkk}
\langle S_{\ell_1}^2\rangle_{\psi} +\langle
S_{\ell_2}^2\rangle_{\psi}+\langle S_{\ell_3}^2\rangle_{\psi}
+\langle S_{\ell_4}^2\rangle_{\psi} \nonumber
\\ + \langle S_{\ell_5}^2\rangle_{\psi} \geq 3
\end{equation}
using the identity $A_iA_{i+1}=2S_{\ell_i}^2+2S_{\ell_{i+1}}^2-3$
easily derived from Eq.~(\ref{Observable}) below. We call it the
{\it pentagram inequality\/}.

\begin{figure}[t]
\vspace{-1cm}
\includegraphics[width=10cm]{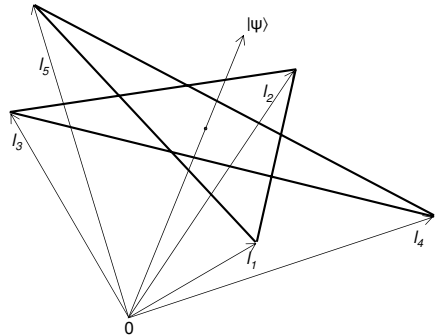}
\vspace{-2.5cm}
\caption{\footnotesize \label{fig1} Regular pentagram defined by cyclic
quintuplet  of unit vectors  $\ell_i \perp \ell_{i+1}$. State
vector $|\psi\rangle$ is directed along the symmetry axis of the
pentagram.}
\end{figure}

Initially the left hand side of the inequality
(\ref{inequality-1}) was found by a computer as an extremal
function of the Vorob'ev--Kellerer cone. The other nontrivial
extremal functions can be obtained from it by flips $a_i\mapsto\pm
a_i$. They, however, add no new physical constraints. For example,
a single flip $A_i\mapsto -A_i$ in Eq.~(\ref{reflect}) yields the
inequality $\langle S_{\ell_i}^2\rangle\le \langle
S_{\ell_{i-2}}^2\rangle+\langle S_{\ell_{i+2}}^2\rangle$. Since in
the pentagram  $\ell_{i-2}\perp\ell_{i+2}$, then $
S_{\ell_{i-2}}^2+S_{\ell_{i+2}}^2+ S_{n_i}^2=2$ for some direction
$n_i\perp\ell_{i\pm2}$, and the inequality becomes  trivial
$\langle S_{\ell_i}^2\rangle +\langle S_{n_i}^2\rangle\le 2$.

In summary, the pentagram inequality, in contrast to the
Kochen-Specker theorem, provides a test for arbitrary hidden
variables model, context-free or not.
Moreover, the inequality is sufficient for existence of such a
model for the observables $S_{\ell_i}^2$. In addition, it reduces
the number of involved spin projection operators from 31, as  in
the best known noncontextual test due to Conway and Kochen, to 5.
As a drawback, the pentagram criterion is state dependent.

A more careful analysis shows that there is no hidden variables test
for three-dimensional quantum system with less then 5 observables.
Furthermore, every such test with 5 observables $A_i$ by an
appropriate scaling $A_i\mapsto \alpha_iA_i+\beta_i$ can be reduced
to the inequality (\ref{reflect}) for a {\it complex\/} pentagram
$\ell_i\perp\ell_{i+1}\in \mathcal{H}$ and
$A_i=1-2|\ell_i\rangle\langle\ell_i|$, cf. Eq.~(\ref{Observable}).
We will provide the details elsewhere.

For further analysis of the pentagram inequality it is convenient
to identify Hilbert space of spin-1 particle with complexification
$\mathcal{H}=\mathbb{E}^3 \otimes \mathbb{C}$ of the physical
Euclidean space $\mathbb{E}^3$. The spin group $\mathrm{SU}(2)$,
locally isomorphic to $\mathrm{SO}(3)$, acts on $\mathcal{H}$ by
rotations in $\mathbb{E}^3$. The cross product $[x,y]=x\times y$
turns Euclidean space $\mathbb{E}^3$ into Lie algebra
$\frak{su}(2)$ and allows to express the spin projection operator
as follows $S_{\ell}\psi=i[\ell,\psi]$. It has three eigenstates,
one real $|0\rangle_\ell=\ell$ and two complex conjugate $|\pm
1\rangle_\ell =(m\pm in)/\sqrt{2}$, where $\{\ell, m, n \}$ is as
an orthonormal basis in $\mathbb{E}^3$. So in this picture the
neutrally polarized spin state $|0\rangle_\ell$ is represented by
{\it real\/} vector $\ell\in\mathbb{E}^3$. The operators $A_i$ are
now given by the equation
\begin{eqnarray}
A_{\ell}=I-2|\ell\rangle\langle\ell|=2S_{\ell}^2-I,
\label{Observable}
\end{eqnarray}
that allows to recast the pentagram inequality into the
geometrical form \begin{eqnarray}
\sum_{k\text{ mod } 5} |\langle\ell_k|\psi\rangle|^2 \leq 2.
\label{Penta-geom}
\end{eqnarray}
Let's test it for a neutrally polarized spin state represented by
a real vector $\psi$ directed along 5-fold symmetry axis of a {\it
regular pentagram\/}, see Fig.~1. A simple calculation shows that
in this case $|\langle\ell_k|\psi\rangle|^2=\cos^2\widehat{\ell_k
\psi}=\frac{1}{\sqrt{5}}$, which violates the pentagram inequality
\begin{eqnarray}
\sum_{k\text{ mod }5}|\langle\ell_k|\psi\rangle|^2=\sqrt{5}
\approx 2.236 >2. \nonumber
\end{eqnarray}
Thus the neutrally polarized spin states are nonclassical. If one
believes in invariance of physical laws with respect to rotations
around $\psi$-axis, then the distributions of spin projections
onto all 5 directions $\ell_k$ of the regular pentagram must be
the same, and only one of them should be actually measured to
refute any hidden variables model. Such a reduction is possible
only for the neutrally polarized spin states, that exhibit the
most extreme nonclassical behavior. One can not achieve that high
symmetry in CHSH setting (\ref{CHSH}), and has to switch the spin
projection directions at both sites which may create a loophole
\cite{review}.

As an example of spin-$1$ system of some  physical interest,
consider $p$-electron in an atom or a molecule with respect to its
{\it orbital\/} momentum, equal to 1, and  disregarding the spin.
In the coherent state $|1\rangle$, with orbital momentum $1$ in
some direction, the electron density looks like a classical Kepler
orbit, while in the neutrally polarized state $|0\rangle$ the
electron splits itself into two blobs separated by a plane of zero
electron density.  In the latter case the electron is hopping
between these two regions never crossing the plane. This state,
known as {\it $p$-orbital\/}, plays a crucial role in chemistry.
Nonclassical nature of this state, and the whole chemistry, can be
detected by the pentagram inequality.

It may be also instructive to look into the meaning of the
pentagram inequality for a composite  spin-1 system formed by two
components $A,B$ of spin-$1/2$. In this setting
$S_\ell=S_\ell^A+S_\ell^B$ and by substitution into
Eq.~(\ref{kkk}) we get a  two-component version of the pentagram
inequality valid for {\it symmetric\/} states of two qubits
\begin{eqnarray}\nonumber
\langle A_1 B_1\rangle+\langle A_2 B_2\rangle+\langle A_3
B_3\rangle+\langle A_4 B_4\rangle+\langle A_5B_5\rangle\geq 1,
\end{eqnarray}
where we use the notations $A_i=2S_{\ell_i}^{A}$,
$B_j=2S_{\ell_j}^{B}$ to facilitate a comparison with CHSH
inequality (\ref{CHSH}). The crucial difference between them is in
the directions of the spin projection measurements at sites $A$
and $B$ which for the pentagram version are always the same. This
allows to detect entanglement in closely tight systems, like atoms
or molecules, where one may not see the separate components. The
latter conclusion holds true even if the components $A,B$ do not
exist outside the system, like quarks or quasiparticles.

These observations may suggest that the nonclassical behavior of
spin-1 system detected by the pentagram inequality originates from
entanglement of its internal degrees of freedom, whatever their
physical nature could be. This is in line with the Majorana
picture of a high spin state as a symmetric state of $2S$ virtual
spin-$1/2$ components readily visualized as a configuration of
$2S$ points in Bloch sphere \cite{Majorana}. A proper name for
this nonclassical effect would be {\it spin state entanglement\/}
\cite{Romeo}.

The above discussion may also clarify  physical meaning  of a more
general notion of ``entanglement beyond subsystems", promoted by
some research groups \cite{Klyachko02,Viola,Kl07}.

Observe that every state of a general spin-$1$ system can be
transformed by a unitary rotation into the {\it canonical form\/}
$$\psi =m \cos \varphi+in \sin\varphi,$$
where $m,n$ are two fixed unit orthogonal vectors in
$\mathbb{E}^3$. Intrinsic properties of $\psi$ are determined by
the parameter $0\le\varphi\le\frac{\pi}{4}$. For example,
Wootters's concurrence $c(\psi)$ \cite{Wootters} of spin state
$\psi$, considered as a symmetric state of two qubits, is equal to
$\cos2\varphi$ and coincides with a measure of entanglement for
spin states introduced in \cite{Can}.  The extremal values $c=0$
and $c=1$ correspond to the coherent $|1\rangle$ and the neutrally
polarized $|0\rangle$ spin states respectively.

Note that a regular pentagram can detect nonclassical nature of a
spin-1 state $\psi$ only for $c(\psi)>\frac{1}{\sqrt{5}}$. For a
state with a smaller positive concurrence one has to use  an
appropriate skew pentagram. On the other hand, coherent states
pass the  test for any pentagram, and hence they are the only
classical spin-1 states. We refer for details to
\cite[$n^\circ$3.4]{Kl07}.

As a convenient physical model of spin-$1$ system suitable for
experimental study consider a single mode {\it biphoton\/}, i.e. a
pair of photons in the same spatiotemporal mode, so that they
differ only in polarization \cite{Biphoton}. The photons obey Bose
statistics, hence their polarization space is spanned by the
symmetric triplet
\begin{eqnarray}
|\circlearrowleft\circlearrowleft\rangle,\quad
\mbox{$\frac{1}{\sqrt{2}}$}(|\circlearrowleft\circlearrowright\rangle+|\circlearrowright\circlearrowleft\rangle),
\quad|\circlearrowright\circlearrowright\rangle,
\label{P-triplet}
\end{eqnarray}
corresponding to spin states $|1\rangle$, $|0\rangle$, $|-1\rangle$.
Here $\circlearrowleft$ and $\circlearrowright$ represent left and
right circularly polarized photons. The biphoton is usually created
via a nonlinear down conversion process in a neutrally polarized
state like the second one in the above triplet.

For the biphoton system the concurrence $c(\psi)$ is closely
related to its {\it degree of polarization\/}
$P(\psi)=\sqrt{1-c(\psi)^2}$, that can be literally seen  in
classical polarization dependent intensity measurements \cite{Bi}.
In contrast, the quantity
$|\langle\ell|\psi\rangle|^2=1-\langle\psi|S_\ell^2|\psi\rangle$
that enters into the pentagram inequality (\ref{Penta-geom})
requires a quantum measurement in a specific setting of the
Hanbury~Brown--Twiss interferometer described below.

The directions $\ell$ for the biphoton should be  taken in the
{\it polarization space\/} $\mathbb{R}^3_{\text{\it pol}}$ with
Stokes parameters $S_1,S_2,S_3$ as coordinates, rather than in the
physical space $\mathbb{E}^3$. The Hilbert state space of a
biphoton is obtained by {\it complexification\/} of the
polarization space. The neutrally polarized states correspond to
{\it real\/} state vectors $\psi\in \mathbb{R}^3_{\text{\it pol}}$
that can be interpreted as follows. Let $P,Q$ be orthogonal
polarization states of a photon described by the antipodal points
$\pm\psi$ of the Poincar\'e sphere
$\mathbb{S}^2\subset\mathbb{R}^3_{\text{\it pol}}$. Then
\begin{equation}
\psi=\mbox{$\frac{1}{\sqrt{2}}$}(|PQ\rangle+|QP\rangle).
\end{equation}

In this setting the quantity $|\langle\ell|\psi\rangle|^2$  is
equal to the coincidence rate in the Hanbury~Brown--Twiss
interferometer feeded by biphotons in state $\psi$ while
polarization filters inserted into its arms select photons in
orthogonal polarization states given by the antipodal points
$\pm\ell$ of the Poincar\'e sphere.

As we have seen above, to test  classical realism for a neutrally
polarized state $\psi$ by a regular pentagram one needs the
coincidence rate  for a single direction $\ell$ such that
$|\langle\ell|\psi\rangle|^2=\cos^2\widehat{\ell\psi}=1/\sqrt{5}$,
which corresponds to the angle
$\delta=\widehat{\ell\psi}\approx0.8383$ radian. Quantum theory
predicts the coincidence rate $1/\sqrt{5}\approx0.4472$, while to
refute hidden variables one needs the rate greater than $0.4$.
However, the available raw experimental data from
\cite[Fig.~8]{Kriv-Kulik} by some reason fall far below the
theoretical curve in a vicinity of $\delta=0.8383$ and provide no
evidence for violation of classical realism in the biphoton
system. The data clearly lack for the required precision.

Recently a nonclassical behavior has been detected in a {\it
local\/} two qubit system formed by a single particle spin and two
of its spatial modes created by a beam splitter \cite{Local_CHSH}.
Since nonlocality here is not an issue, the authors interpret the
result as a test of noncontexuality. This may be an
underestimation: a violation of CHSH inequality refutes any hidden
parameters, context-free or not \cite{Fine}.

In conclusion, we close  the gap between two-dimen\-sional quantum
systems, admitting hidden variables description \cite{Bell}, and
four-dimensional systems that are incompatible with such a model
\cite{Fine} by constructing a Bell's type inequality for
three-dimensional spin-$1$ system. We shortly discuss some
physical implications and an underlying cause of the nonclassical
behavior in spin-$1$ systems, as well as a perspective of its
experimental verification.


This work was partially supported by Institute of Materials
Science and Nanotechnology (UNAM), Institute of Theoretical and
Applied Physics (ITAP), and T\"{U}B{\.I}TAK.


\begin{thebibliography}{99}
\bibitem{Romeo} What's in a name? That which we call a rose \\ By any other name would smell as
sweet.\\ (Shakespeare, Romeo and Juliet, II, ii, 1-2)
\end{thebibliography}
\end{document}